RESEARCH ARTICLE | OCTOBER 18 2024

# A machine learning approach to automation and uncertainty evaluation for self-validating thermocouples ⊘

S. Bilson ✉; A. Thompson; D. Tucker; J. Pearce

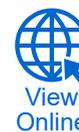
View
Online

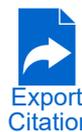
Export
Citation

# A Machine Learning Approach to Automation and Uncertainty Evaluation for Self-Validating Thermocouples

S. Bilson[1, a)], A. Thompson[1], D. Tucker[1] and J. Pearce[1]

[1]*National Physical Laboratory, Hampton Road, Teddington, United Kingdom TW11 0LW*

a)Corresponding author: sam.bilson@npl.co.uk

**Abstract.** Thermocouples are in widespread use in industry, but they are particularly susceptible to calibration drift in harsh environments. Self-validating thermocouples aim to address this issue by using a miniature phase-change cell (fixed-point) in close proximity to the measurement junction (tip) of the thermocouple. The fixed point is a crucible containing an ingot of metal with a known melting temperature. When the process temperature being monitored passes through the melting temperature of the ingot, the thermocouple output exhibits a "plateau" during melting. Since the melting temperature of the ingot is known, the thermocouple can be recalibrated *in situ*. Identifying the melting plateau to determine the onset of melting is reasonably well established but requires manual intervention involving zooming in on the region around the actual melting temperature, a process which can depend on the shape of the melting plateau. For the first time, we present a novel machine learning approach to recognize and identify the characteristic shape of the melting plateau and once identified, to quantify the point at which melting begins, along with its associated uncertainty. This removes the need for human intervention in locating and characterizing the melting point. Results from test data provided by CCPI Europe show 100% accuracy of melting plateau detection. They also show a cross-validated R$^2$ of 0.99 on predictions of calibration drift.

## INTRODUCTION

Most thermometry makes use of practical sensors such as thermocouples. These yield a temperature dependent property such as voltage, which must then be related to temperature by comparison with a set of known temperatures, i.e., a calibration. Thermocouples are in widespread use in industry, but they are particularly susceptible to calibration drift in harsh environments (e.g., high temperatures, contamination, vibration, ionizing radiation) which degrade the thermocouple wire. The amount by which the drift in the relationship between temperature and emf (which shall be called the drift) is generally unknown, which leads to loss of information on the process temperature. This drift can now be monitored *in situ* by using a miniature phase-change cell (fixed point) in close proximity to the measurement junction (tip) of the thermocouple. The fixed point is a crucible containing an ingot of metal (or metal-carbon alloy or organic material) with a known melting temperature. The latest devices developed by NPL's Temperature & Humidity group, in collaboration with thermocouple manufacturer CCPI Europe Limited, are able to accommodate the entire thermocouple and fixed-point assembly within a protective sheath of outer diameter 7 mm. Importantly, this means that the self-validating thermocouple presents the same external form factor and appearance as a regular process control thermocouple. A self-validating thermocouple is shown in Fig. 1.

In use, when the process temperature being monitored passes through the melting temperature of the ingot, the thermocouple output exhibits a "plateau" during melting, due to the heat of fusion of the ingot restraining further temperature rise. Once the ingot is completely melted, the indicated temperature resumes its upward trend. As the melting temperature of the ingot is known, the thermocouple can be recalibrated *in situ*. A typical melt and freeze cycle of a self-validating thermocouple during the recalibration process is shown in Fig. 2.

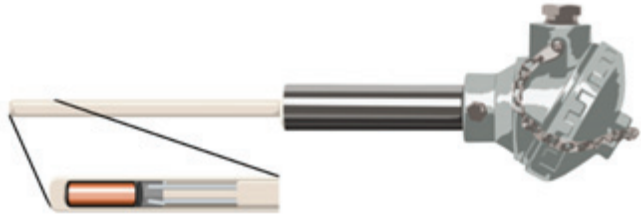

**FIGURE 1.** Self-validating thermocouple with protective sheath. Image courtesy of CCPI Europe.

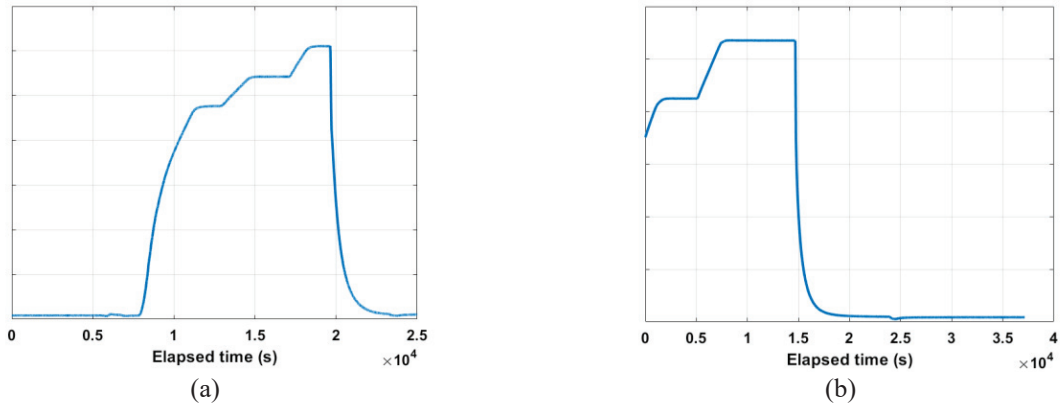

**FIGURE 2.** Example (a) silver and (b) gold cycles. The temperature-time data is proprietary commercial information, so the ordinate axis label has been removed.

This device has been extensively characterized by CCPI Europe under the trade name INSEVA, who are executing a series of trials in high value manufacturing industries. The NPL/CCPI self-validating thermocouple has been demonstrated in several industrial heat treatment applications and has been shown to be robust and fit for purpose, in that clear melting plateaus are realized which permit reliable *in-situ* recalibration. The main outstanding problem now is to automate the identification and characterization of the melting plateau (see Fig. 3).

The process of identifying the melting plateau to determine the onset of melting (the invariant part of the melting plateau, i.e., the part that is least dependent on ambient thermal influences) is reasonably well established but requires manual intervention to identify the region around the actual melting temperature and then characterize it; for this, the technique to be used depends on the shape of the melting plateau. Attempts to develop algorithms to do this using conventional techniques (using numeric measures of the first derivative of indicated temperature with respect to time) have not been successful, due to the existence of additional and sometimes unpredictable gradient changes around the melting plateau. For the first time, a novel machine learning approach was developed to identify the characteristic shape of the melting plateau and once identified, to estimate the point at which melting begins.

Data in the form of temperature indicated by the thermocouple as a function of time, obtained from industrial trials of self-validating type S thermocouples, was used for training and testing the machine learning algorithms. The self-validating thermocouples used either silver or gold ingots, with melting points of 961.78 °C and 1064.18 °C respectively.

## METHODS

### Detecting the Melting Plateau

The thermocouple will undergo thermal cycles in operation when the same process is carried out repeatedly. Cycles can consist of multiple changes in ramp rate (i.e., rate of change of temperature with time), combined with long periods where the melting plateau ramp rate changes are on a smaller scale (see Fig. 2). This can be a challenge for any detection algorithm since the melting plateau might be obscured within a larger change in ramp rate, as is the case for gold melts, where the melting temperature is close to the maximum temperature reached during the industrial process. In this section, we propose a method based upon a piecewise model which can detect the smaller changes in temperature gradient through supervised learning.

The piecewise model used in this work is based upon a parameter $\lambda \in [0, \infty)$, which controls the number of changepoints detected. i.e. the number of clear changes in gradient in the time-series. This number may change for each cycle for a fixed $\lambda$, however the aim is to have enough changepoints to capture the behavior of the melting plateau within the cycle, but not too many to obscure it (see Fig. 3). To determine the number of changepoints required in each cycle for melting plateau detection, specific, characteristic behavior of the melting plateau must be satisfied. This was broken down into criteria through discussion with domain experts. If there is no changepoint that meets all the criteria, then we state that no melting plateau exists within the data.

To optimize for $\lambda$ such that melting plateaus are detected, a supervised learning approach was taken where the existence or absence of a melting plateau was known from manual inspection. Melting plateaus do not exist when the temperature of the furnace does not reach the melting temperature of the metal. If we assign to each of $N$ cycles a ground truth label ($c_i = 1$ if melting plateau exists, and $c_i = 0$ otherwise), the task can be framed as one of binary classification. If we let the predicted label be $\hat{c}_i(\lambda)$, then $\lambda$ is found by minimizing the classification loss given in (1).

$$Loss(\lambda) = \frac{1}{N} \sum_{i=1}^{N} |c_i - \hat{c}_i(\lambda)|$$ (1)

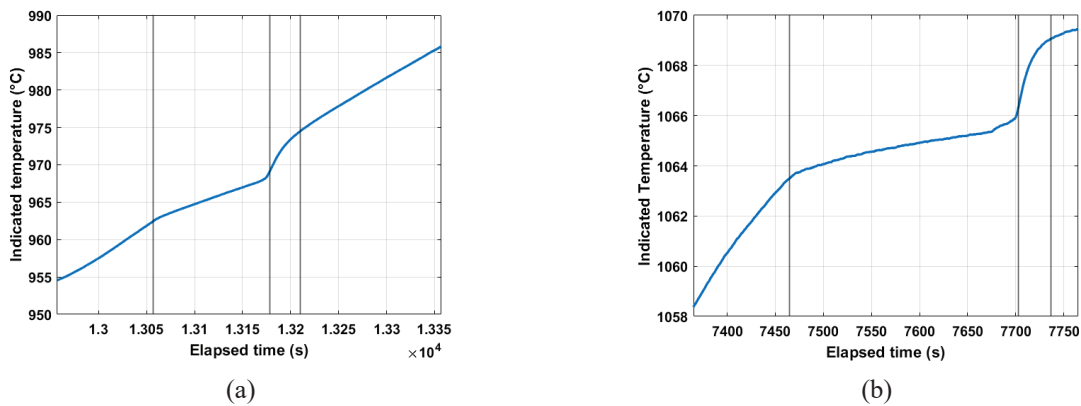

(a)                                                     (b)

**FIGURE 3.** Optimal number of changepoints (indicated by the vertical lines) for silver (a) and gold (b) melting plateaus from the cycles shown in Fig. 2.

To perform the optimization, a Bayesian optimization procedure (using *bayesopt* in MATLAB R2022a) was chosen on training data consisting of 140 cycles with known melting plateaus (melts) and 98 non-melts, with 50 observed points. The optimization procedure consists of a Gaussian process model (GPR) of the loss (1). We see from Fig. 4(a) that there is a region of minimum loss where one can extract the bounds $\lambda_{low}$ and $\lambda_{high}$. If $\lambda$ is too small, there are too many changepoints which start to capture noise in the time-series, and therefore melting plateau is obscured. However, if $\lambda$ is too large, there are not enough changepoints to characterize the plateau.

## Estimating the Onset of Melting

Once we have a model for detecting the melting plateau, the next task is to estimate the onset of melting as measured by the thermocouple. To do this, one must come up with an appropriate model for the abrupt phase change taking place, from which the temperature at this changepoint can be extracted. A segmented quadratic-quadratic model is proposed within a truncated region around the critical changepoint, using a truncation parameter $\gamma \in (0,1]$. When inspecting the residuals in the model fits, there was clear autocorrelation, and so an autoregressive AR(1) model was applied to the noise. Given the truncated temperature data $\mathcal{D} = \{t_i, y_i\}$ where $i = n_1, ..., n_2$, the model given in (2) has nine parameters $\theta = (m, \beta_1, ..., \beta_6, \phi, \sigma)$, where $m$ is the changepoint parameter, $\{\beta_j\}$ are the quadratic regression coefficients, and $(\phi, \sigma)$ are the AR(1) lag and noise parameters respectively.

$$y_i = \begin{cases} \beta_1 + \beta_2 t_i + \beta_3 t_i^2 + \phi y_{i-1} + \epsilon_i & n_1 \le i \le m-1 \\ \beta_4 + \beta_5 t_i + \beta_6 t_i^2 + \phi y_{i-1} + \epsilon_i & m \le i \le n_2 \end{cases} \tag{2}$$

To calculate the onset of melting predictions, we take a Bayesian inference approach to calculate the marginal posteriors of the model parameters. Given the likelihood $\mathcal{L}(\mathcal{D}|\theta)$ of model (2), and an uninformative Jeffrey's prior $\pi(\theta)$, the posterior $p(\theta|\mathcal{D})$ can be determined using Bayes' rule:

$$p(\theta|\mathcal{D}) = \frac{\mathcal{L}(\mathcal{D}|\theta)\pi(\theta)}{p(\mathcal{D})} \propto \mathcal{L}(\mathcal{D}|\theta)\pi(\theta) \tag{3}$$

Estimation of $\theta$ involves calculating the maximum *a posteriori* estimate (MAPE), which is a global optimization problem. Optimization was performed by minimizing the negative log posterior using a sequential quadratic programming algorithm (see Fig. 4(a)). Once the MAPEs are known, then the onset of melting can be determined as the crossing point of the two quadratic functions as in Fig. 5.

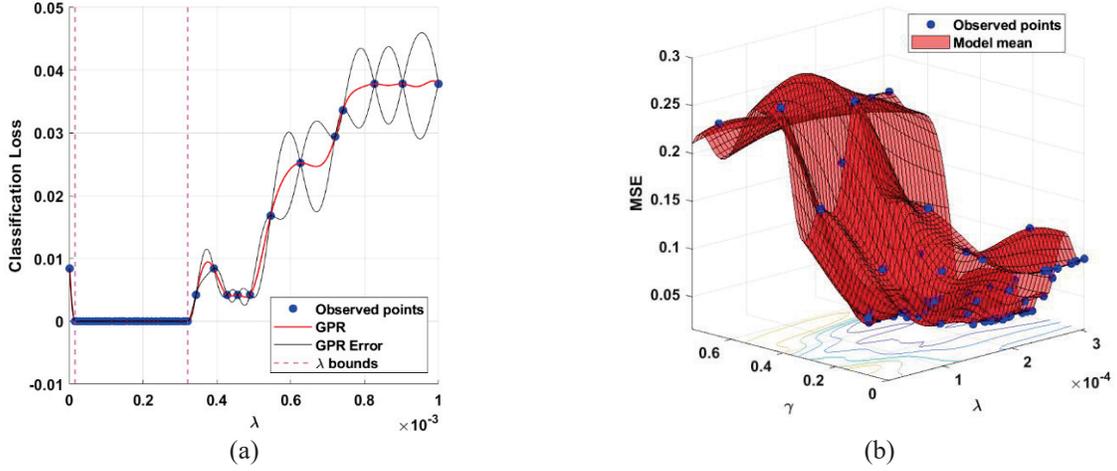

**FIGURE 4.** Bayesian optimization of model parameters $(\lambda, \gamma)$. (a) shows the identified bounds $\lambda_{low}$ and $\lambda_{high}$ where the classification loss (1) is minimized. (b) uses those bounds to minimize the MSE (4) in both parameters.

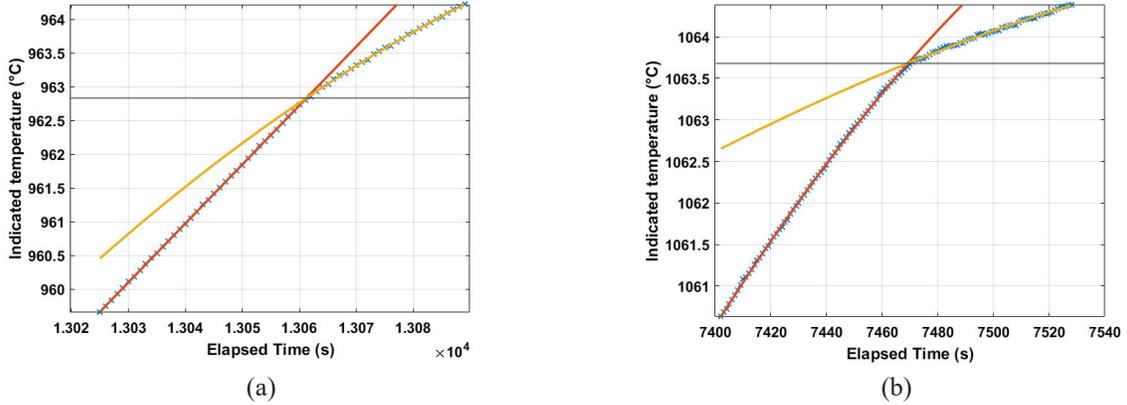

**FIGURE 5.** Examples of predicting the onset of melting as indicated by a thermocouple with (a) silver or (b) gold fixed-point, using the MAPEs for model (2) with quadratic fits and onset of melting estimates shown.

We use a supervised learning approach to determine estimates of the machine learning parameters $(\lambda, \gamma)$. Ground truth for onset of melting $T_i$ were recorded via manual visual inspection of the changepoint (see Fig. 7(a) for the ground truth data). Given the predicted onset of melting $\hat{T}(\lambda, \gamma)$, this can be framed as a regression problem where we need to minimize the mean squared error (MSE) given in (4).

$$MSE(\lambda, \gamma) = \frac{1}{N} \sum_{i=1}^{N} \left( T_i - \hat{T}_i(\lambda, \gamma) \right)^2 \qquad (4)$$

We use the same Bayesian optimization procedure already described (see Fig. 4(b)) to minimize the MSE given in (4) over 100 observed points, using the parameter bounds $\lambda \in (\lambda_{min}, \lambda_{max})$ and $\gamma \in (0,1]$. The final estimates $(\hat{\lambda}, \hat{\gamma})$ are determined using the minimum of the GPR fit to the observed points.

The uncertainty in the onset of melting predictions is derived from the posterior distribution of the model parameters given in (3). However, the denominator in (3) is analytically intractable, so sampling methods were deployed. More specifically we use a Metropolis-Hastings Markov Chain Monte Carlo (MCMC) sampling algorithm for each melting plateau (performed in MATLAB R2022a). To generate marginal posterior distributions for the parameters in (2), we take $10^6$ samples from four chains (see Fig. 6(a)). The posteriors for the regression coefficients are then used to determine the distribution in the estimated onset of melting (see Fig. 6(b)).

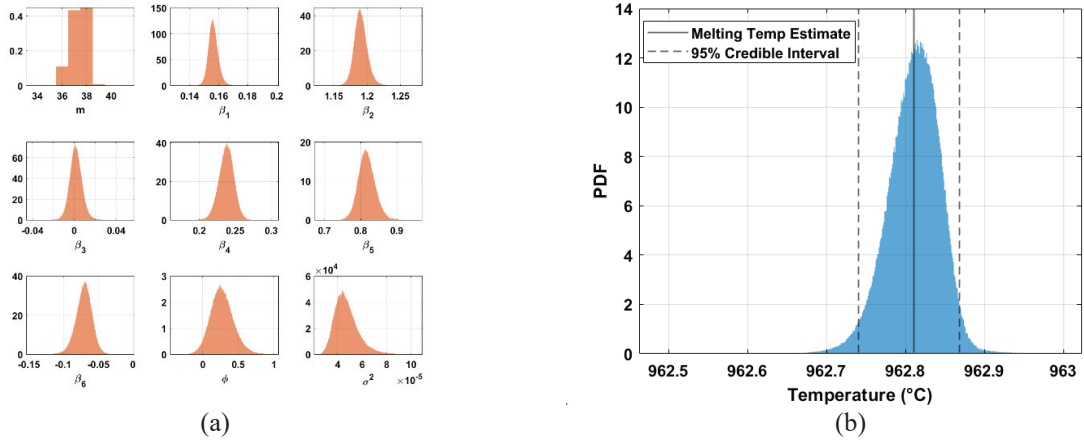

(a)                                         (b)

**FIGURE 6.** (a) Marginal posteriors of model (2) using MCMC on (3) for a silver melting plateau; (b) the resulting distribution in estimated onset of melting, with a 95% credible interval added to quantify the uncertainty in the model prediction.

## RESULTS

The methods were applied to both a training set of 238 cycles, which consisted of 140 melting plateaus (melts) and 98 non-melts, and an "unseen" test set, which was provided after the model was trained, consisting of 121 cycles, including 79 cycles with melts, and 42 non-melts. The training and test sets had a mixture of both silver and gold melts.

The model has a 99.5 % 5-fold cross-validation training accuracy of melting plateau detection, and a 100 % accuracy on the test set. The model also has a 5-fold cross-validated $R^2$ of 0.9927 on predicting drift. The results of the predictions along with the 95 % credible intervals are given in Fig. 7.

## CONCLUSION

In this paper, we provide a two-parameter machine learning model for identifying the melting plateau and estimating the drift in melting temperature of metal ingots in self-validating thermocouples. The results show extremely good performance of melt detection (99.5 % and 100 % accuracy on cross-validated training and test sets respectively) and calibration drift estimation (cross-validated $R^2$ of 0.9927). The models trained can be generalized to self-validating thermocouples with different metal ingots, by just specifying the actual metal melting temperature. The performance of the machine learning model has been demonstrated to be sufficiently good to enable autonomous operation of the self-validating thermocouples for the data provided so far.

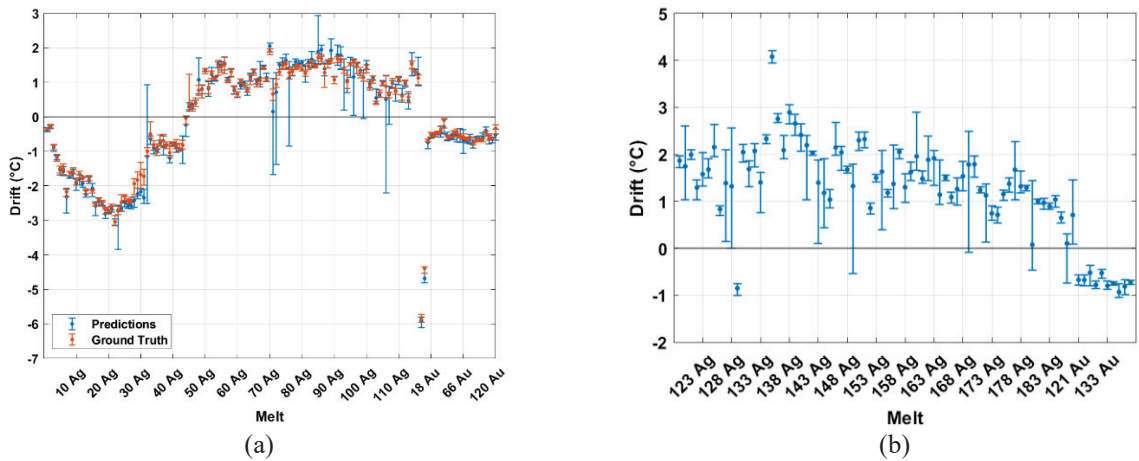

**FIGURE 7**. Predictions of calibration drift (the difference between the estimate of onset of melting and the actual melting temperature) on (a) training data, where the ground truth is included for comparison, and (b) test data.

## ACKNOWLEDGMENTS

We would like to thank Phill Williams, Peter Cowley, and Trevor Ford (CCPI Europe) for conducting the trials and providing the data. We also thank Peter Harris (NPL) for useful guidance, and Miles McCrory (University of Surrey) for labelling the indicated melting temperatures. This work was supported by the UK Government's Department for Science, Innovation and Technology (DSIT).